\documentclass{kapproc} 

\RequirePackage{graphicx}%
\RequirePackage{epsf}%
\input{psfig.sty}

\upperandlowercase
\let\footnote\savefootnote
\let\footnotetext\savefootnotetext 
\let\footnoterule\savefootnoterule 
\kluwerbib 
\newcommand{\oversim}[2]{\protect{\mbox{\lower0.5ex\vbox{%
   \baselineskip=0pt\lineskip=0.2ex
   \ialign{$\mathsurround=0pt #1\hfil##\hfil$\crcr#2\crcr\sim\crcr}}}}}
\newcommand{\simgreat}{\mbox{$\,\mathrel{\mathpalette\oversim>}\,$}} 
\newcommand{\simless} {\mbox{$\,\mathrel{\mathpalette\oversim<}\,$}} 
\begin{document}


\articletitle{Variations of the IMF}


\chaptitlerunninghead{Variation of the IMF}\footnotetext{to appear in {\it
    IMF@50: The Initial Mass Function 50 years later}, ed: E.
  Corbelli, F. Palla, and H. Zinnecker, Kluwer Academic Publishers; a
  meeting held at the Abbazia di Spineto, Tuscany, Italy -- May 16-20,
  2004}



 \author{Pavel Kroupa and Carsten Weidner}
 \affil{Sternwarte, University of Bonn, Auf dem H\"ugel 71,
 D-53121 Bonn, Germany}
 \email{pavel@astro.uni-bonn.de, cweidner@astro.uni-bonn.de}





\begin{abstract}
  The {\it stellar IMF} has been found to be essentially invariant.
  That means, any detailed observational scrutiny of well resolved
  populations of very young stars appears to show the IMF to be of
  more or less the same shape independent of the metallicity and
  density of the star-forming region. While some apparent differences
  are seen, the uncertainties inherent to this game do not allow a
  firm conclusion to be made that the IMF varies systematically with
  conditions. The IMF integrated over entire galaxies, however, is
  another matter. Chemical and photometric properties of various
  galaxies do hint at {\it galaxial IMFs} being steeper than the
  stellar IMF, as is also deduced from direct star-count analysis in
  the MW. These results are however sensitive to the modelling of
  stellar populations and to corrections for stellar evolution, and
  are thus also uncertain. However, by realising that galaxies are
  made from dissolving star clusters, star clusters being viewed as
  {\it the fundamental building blocks of galaxies}, the result is
  found that galaxial IMFs must be significantly steeper than the
  stellar IMF, because the former results from a folding of the latter
  with the star-cluster mass function. Furthermore, this notion leads
  to the important insight that galaxial IMFs must vary with galaxy
  mass, and that the galaxial IMF is a strongly varying function of
  the star-formation history for galaxies that have assembled only a
  small mass in stars. Cosmological implications of this are that the
  number of SNII per low-mass star is significantly depressed and that
  chemical enrichment proceeds much slower in all types of galaxies,
  and particularly slowly in galaxies with a low average
  star-formation rate over what is expected for an invariant Salpeter
  IMF. Using an invariant Salpeter IMF also leads to wrong $M/L$
  ratios for galaxies. The detailed implications need to be studied in
  the future.
\end{abstract}


\section{The shape of the stellar IMF}
\label{sec:shape}

\noindent
The number of stars in the mass interval $m,~m+dm$ is $dN \equiv
\xi(m)\,dm$, where $\xi(m)$ is the stellar initial mass function
(IMF).  The logarithmic slope is $\Gamma(\log_{10}m) =
{d\log_{10}\xi_{\rm L}(\log_{10}m)}/{d\log_{10}m}$ such that
$\log_{10}\xi(m) = \log_{10}k - \alpha \log_{10}m$ for the power-law
form, $\xi(m) = k m^{-\alpha}$, and $\alpha(m) = -d\log_{10}\xi(m)/
d\log_{10}m = 1 - \Gamma(m)$.  The {\it logarithmic IMF} is $\xi_{\rm
  L}(m) = m\,{\rm ln}(10)\,\xi(m)$, which is useful when counting
stars in logarithmic mass intervals.

The {\it stellar IMF} is the distribution of stellar masses that
results from a single star-formation burst. Because essentially all
stars form in clusters (Lada \& Lada 2003) we expect to measure the
stellar IMF in young star clusters before they disperse to the field
of a galaxy.

The stellar IMF is one of the most fundamental distribution functions
of astrophysics, and consequently a huge effort has been invested into
constraining its shape since its first formulation in 1954 by Salpeter
(1955) (at the Australian National University in Canberra) as a single
power-law with $\alpha=2.35$ for $0.4<m/M_\odot < 10$, based on an
early analysis of star-counts in the solar neighbourhood.  A further
milestone in this remarkable scientific enterprise is given by Miller
\& Scalo (1979) who made a large effort in constraining the IMF to
mass ranges outside the Salpeter limits and who deduced that the
stellar IMF flattens below $0.5\,M_\odot$.  Scalo (1986) re-considered
this problem in what today remains the most significant piece of work
in this topic by studying all of the then available observational
constraints on local star counts, the shape of the stellar luminosity
function, the mass-luminosity relation, stellar evolution for
early-type stars and the vertical structure of the Milky Way (MW), and
suggested the IMF to turn-down below about $0.4\,M_\odot$, which had
important implications in trying to understand the nature and
occurrence of unseen matter in the disk of the MW.  An important
improvement in understanding the shape of the IMF for low-mass stars
was contributed by Kroupa, Tout \& Gilmore (1993 and two prior papers)
who proposed that the strong maximum near $M_{\rm V}\approx 12$ in the
luminosity function of solar-neighbourhood stars stems from an
inflection in the mass--luminosity relation, which in turn results
from the association of H$_2$ molecules and the onset of full
convection below about $0.4\,M_\odot$. This revised the IMF to be
essentially flat below $0.4\,M_\odot$, but the inclusion of
corrections for unresolved multiple systems and detailed modelling of
star-counts with Malmquist bias and galactic-disk structure solved the
disagreement between local and deep star counts and thereby increased
the slope of the IMF below $0.5\,M_\odot$.  The resulting form of the
two-part power-law IMF for late-type stars ($\alpha_1\approx 1.3:
0.08\simless m/M_\odot \simless 0.5; \alpha_2\approx 2.3: 0.5\simless
m/M_\odot\simless 1$) has been verified by Reid, Gizis \& Hawley
(2002) using revised local star-counts that incorporate {\sc
  Hipparcos} distances. For $m\simgreat 1\,M_\odot$ the IMF is
proposed by Reid et al. to have $\alpha_3=2.5-2.8$, while Scalo (1986)
derived $\alpha_3\approx 2.7$, which is supported by the work of Yuan
(1992).  Written out,

\vspace{-5mm}

{\small
\begin{equation}
\xi(m) = k \left\{\begin{array}{ll}
\left(\frac{m}{m_{\rm H}} \right)^{-\alpha_{0}}&\hspace{-0.25cm},m_{\rm
  low} \le m < m_{\rm H},\\
\left(\frac{m}{m_{\rm H}} \right)^{-\alpha_{1}}&\hspace{-0.25cm},m_{\rm
  H} \le m < m_{0},\\
\left(\frac{m_{0}}{m_{\rm H}} \right)^{-\alpha_{1}}
  \left(\frac{m}{m_{0}} \right)^{-\alpha_{2}}&\hspace{-0.25cm},m_{0}
  \le m < m_{1},\\ 
\left(\frac{m_{0}}{m_{\rm H}} \right)^{-\alpha_{1}}
    \left(\frac{m_{1}}{m_{0}} \right)^{-\alpha_{2}}
    \left(\frac{m}{m_{1}} \right)^{-\alpha_{3}}&\hspace{-0.25cm},m_{1}
    \le m < m_{\rm max},\\ 
\end{array} \right. 
\label{eq:4pow}
\end{equation}
\noindent with exponents
\begin{equation}
          \begin{array}{l@{\quad\quad,\quad}l}
\alpha_0 = +0.3\pm 0.7&0.01 \le m/{M}_\odot < 0.08,\\
\alpha_1 = +1.3\pm 0.5&0.08 \le m/{M}_\odot < 0.50,\\
\alpha_2 = +2.3\pm 0.3&0.50 \le m/{M}_\odot < 1.00,\\
\alpha_3 = +2.7\pm 0.7&1.00 \le m/{M}_\odot,\\
          \end{array}
\label{eq:imf}
\end{equation}}
is the KTG93 IMF after extension to the sub-stellar mass range (Kroupa
2001; 2002). This IMF is the {\it galaxial IMF} as it is derived from
galactic field stars, and, as we will see further below, differs from
the {\it stellar IMF}.  The multi-power-law description is used merely
for convenience, as it allows us to leave the low-mass part of the IMF
unchanged while experimenting with different slopes above, say,
$1\,M_\odot$.  Other functional forms have been suggested (Miller \&
Scalo 1978; Larson 1998; Chabrier 2001), but these bear the
disadvantage that the whole functional form reacts to changes in the
parameters. Such parametrisations are useful for studying possible
changes of the stellar IMF with cosmological epoch.

To constrain the stellar IMF we need detailed observations of young
populations in star clusters and OB associations. The hope is that
young clusters are not yet dynamically evolved so that the initial
stellar population becomes evident. However, as dynamical modelling of
young clusters containing O stars shows, significant dynamical
evolution is already well established at an age of 1~Myr as a result
of (i) the dense conditions before gas removal such that the
binary-star population changes significantly away from its primordial
properties, and (ii) rapid cluster expansion as a result of violent
gas expulsion (Kroupa, Aarseth \& Hurley 2001). This would lead to a
systematic biasing of the observed stellar IMF against low-mass
members if the cluster was significantly mass-segregated prior to gas
expulsion (Moraux, Kroupa \& Bouvier 2004). In addition to these
difficulties come obscuration by natal gas and dust and the large
uncertainties in deriving ages and masses given inadequate theoretical
models of stars that are still relaxing from their accretion history,
rotating rapidly, are variable and active, and have circum-stellar
material. 

The task of inferring the stellar IMF is thus terribly involved and
uncertain. Nevertheless, substantial progress has been achieved.
Given that there are many star clusters and OB associations, a large
body of literature has been amassing over the decades, and Scalo
(1998) compiled the then available IMF slope vs mass data. 

\begin{figure}[ht]
\sidebyside
{\centerline{\psfig{file=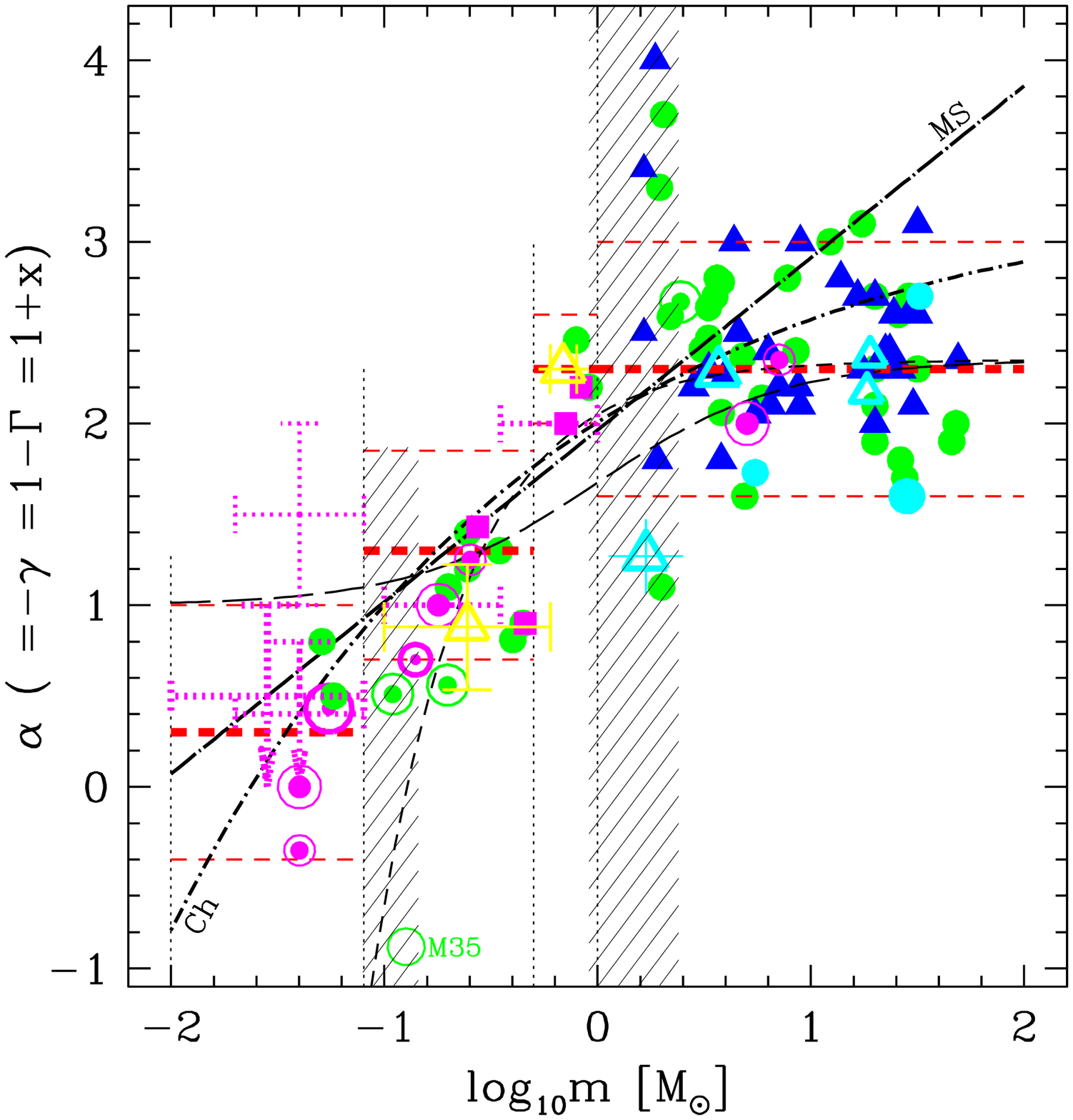,width=7.5cm}}
\vspace*{-2cm}
\caption{The alpha plot compiles measurements of the power-law index,
  $\alpha$, as a function of the logarithmic stellar mass and so
  measures the shape of the MF. The canonical IMF
  (eq.~\ref{eq:canonicalIMF}) is represented by the thick short-dashed
  horizontal lines, and other functional forms are shown using other
  line-types (MS indicates the Miller \& Scalo (1979) log-normal form;
  and the thin short- and long-dashed lines are from Larson (1998),
  while the thick short-dash-dotted curve is the Chabrier (2001) form
  labelled by Ch). Shaded regions indicate mass-ranges over which
  derivation of the MF is particularly hard.  For details see Kroupa
  (2002).  }
\label{krFig1}}
{\centerline{\psfig{file=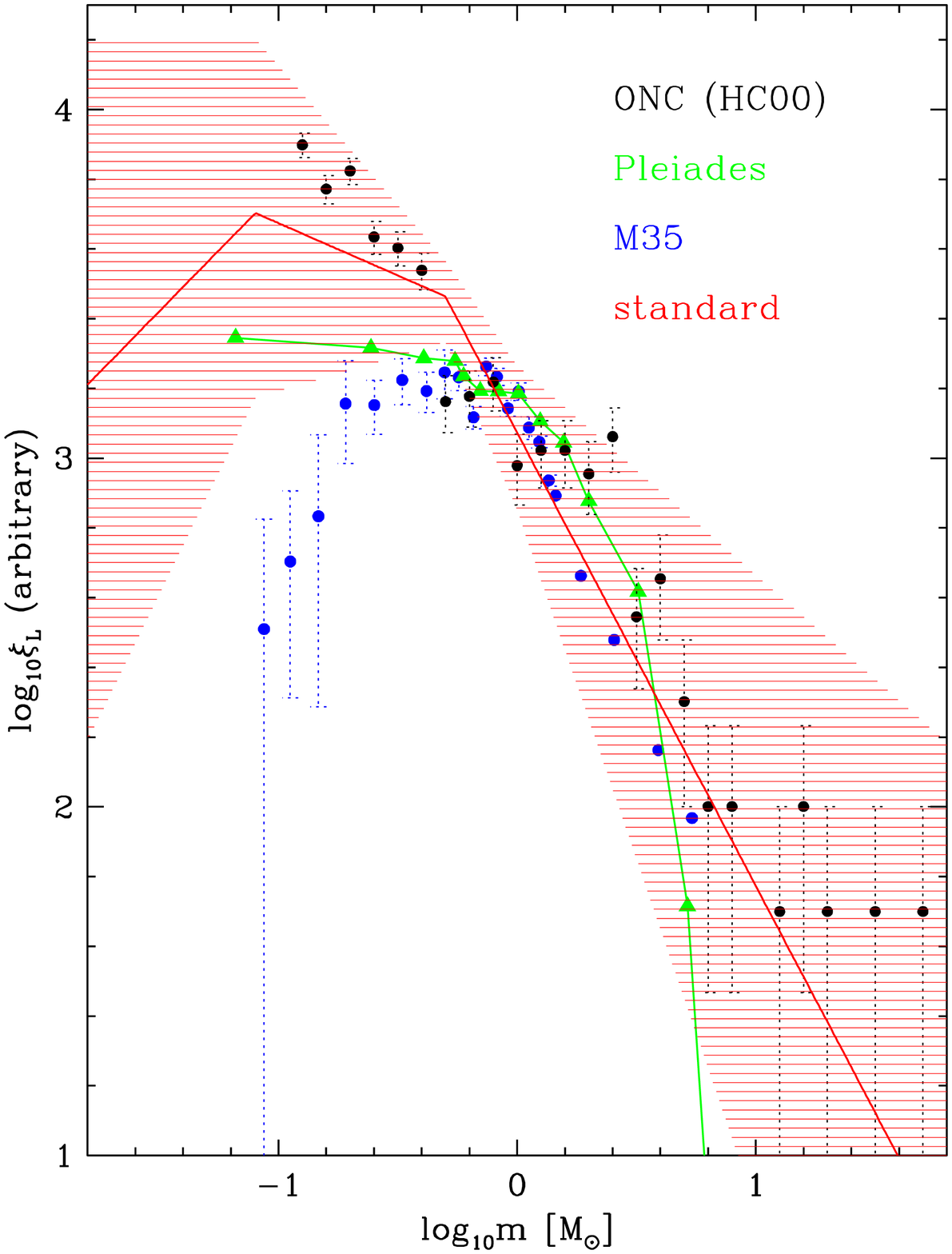,width=7.0cm}}
\vspace*{-0.5cm}
\caption{The measured logarithmic stellar mass functions, $\xi_{\rm L}$, 
  in the Orion Nebula Cluster [ONC, solid circles], the Pleiades
  [triangles] and the cluster M35 [lower solid circles]. The average
  Galactic field single star-star IMF (eq.~\ref{eq:imf}) is shown as
  the solid line with the associated uncertainty range. For details
  see Kroupa (2002).}
\label{krFig2}}
\end{figure}

An updated form of this plot is shown in Fig.~\ref{krFig1}, and as can
be seen, the stellar IMF essentially follows the above galaxial IMF
below $1\,M_\odot$ (after correcting the observations for unresolved
multiple systems), whilst having a Salpeter slope above $1\,M_\odot$.
Striking is that Large and Small Magellanic Cloud data (solid
triangles) are not systematically different to the more metal rich MW
data (solid circles). A systematic difference between dense clusters
and sparse OB associations is also not evident (Massey 1998).  Notable
is that the scatter is large above $1\,M_\odot$ but constant and that
it can be fitted by a Gauss distribution of $\alpha$ values centred on
the Salpeter index (Kroupa 2002). The scatter can be understood to
result from statistical fluctuations and stellar-dynamical evolution
of the clusters and the dynamically evolved OB associations (Elmegreen
1999; Kroupa 2001).  These observations on the $\alpha$-plot thus
point to a remarkable uniformity of the stellar IMF, which can thus be
summarised by the multi-power-law form of eq.~\ref{eq:imf} but with
\begin{equation}
\alpha_3=2.3.
\label{eq:canonicalIMF}
\end{equation}
Some apparently significant deviations from this {\it standard} or
{\it canonical} IMF do occur though (Fig.~\ref{krFig2}), and it will be
the aim of observers and modellers alike to try to understand why the
two clusters of similar age, the Pleiades (Hambly et al. 1999) and M35
(Barrado y Navascu\'es et al. 2001), appear to have such different mass
functions below $0.5\,M_\odot$.

The existence of a universal canonical stellar IMF is therefore
supported by the majority of data. It has a Salpeter index above about
$0.5\,M_\odot$.  But this result uncovers a possibly unsettling
discrepancy between this stellar IMF (eq.~\ref{eq:canonicalIMF}) and
the galaxial IMF (eq.~\ref{eq:imf}) which was found to be steeper for
$m>1\,M_\odot$.

\section{The galaxial IMF}
\label{sec:field}

\noindent
The distribution of stellar masses in an entire galaxy results from
the addition of all star-forming events ever to have occurred,
\begin{equation} 
\label{eq:igimf}
\xi_{\rm IGIMF}(m) = \int_{M_{\rm ecl,min}}^{M_{\rm ecl,max}}
\xi(m\le m_{\rm max})~\xi_{\rm ecl}(M_{\rm ecl})~dM_{\rm ecl}.
\end{equation}
This is the integrated {\it galaxial IMF} (IGIMF), i.e. the IMF
integrated over space and time.  Here $\xi_{\rm ecl}$ is the MF of
embedded clusters, and $M_{\rm ecl,min}=5\,M_\odot, M_{\rm
  ecl,max}\simless 10^6\,M_\odot$ are the minimum and maximum cluster
masses, respectively. Note that the IGIMF becomes indistinguishable
to the field-star IMF in galaxies in which presently on-going
star-formation contributes insignificantly to the already present
stellar population, and also that $\xi_{\rm IGIMF}$ does not
correspond to the shining matter distribution for $m\simgreat
1\,M_\odot$. To evaluate this we need to consider only the recently
formed stars.

Within each cluster stars are formed following the canonical IMF.
However, small star-forming cloud cores, similar to the individual
groups containing a dozen $M_\odot$ in gas seen in Taurus--Auriga, for
example, can never from O~stars. Since the stellar IMF has been found
to be essentially invariant, even when comparing the small groups of
dozens of pre-main sequence stars in Taurus--Auriga with the rich
Orion Nebula Cluster (Kroupa et al.  2003), constructing a young
cluster population is mathematically equivalent to randomly sampling
the canonical IMF. However, sampling without a mass constraint would
in principle allow small star-forming regions to form very massive
stars, which would be in violation to the observations according to
which massive stars are formed in rich clusters.  We therefore choose
stars randomly from the canonical IMF but impose a mass-limit on the
mass in stars that can form from a molecular cloud core with mass
$M_{\rm core}=M_{\rm ecl}/\epsilon$, where $\epsilon\simless 0.4$ is
the star-formation efficiency (Lada \& Lada 2003),
\begin{equation}
\label{eq:Mecl}
M_{\rm ecl} = \int_{m_{\rm low}}^{m_{\rm max}}m \cdot \xi(m)~dm,
\end{equation}
where $m_{\rm low}=0.01\,M_\odot$.  In each cluster there is one
single most-massive star with mass $m_{\rm max}$,
\begin{equation}
\label{eq:normlim}
1 = \int_{m_{\rm max}}^{m_{\rm max *}}\xi(m)~dm,
\end{equation}
where $m_{\rm max*}\approx150\,M_\odot$ is the fundamental upper
stellar mass limit (Weidner \& Kroupa 2004a). This set of two
equations needs to be solved numerically to obtain $m_{\rm max} =
m_{\rm max}(M_{\rm ecl})$ that enters eq.~\ref{eq:igimf}.  We note
that our usage of an $m_{\rm max}(M_{\rm ecl})$ correlation disagrees
with Elmegreen's (2004) conjecture that no such relation exists. This
problem is dealt with further by Weidner \& Kroupa (2004b).

The slope, $\beta$, of the star cluster IMF, $\xi_{\rm ecl} \propto
M_{\rm ecl}^{-\beta}$, is constrained by observations:
\begin{itemize}
\item $20 \le M_{\rm ecl}/M_{\odot} \le 1100$: $\beta \approx 2$
  locally (Lada \& Lada 2003);
\item $10^{3} \le M_{\rm ecl}/M_{\odot} \le 10^{4}$: $\beta \approx
  2.2$ for LMC and SMC (Hunter et al. 2003)
\item $10^{4} \le M_{\rm ecl}/M_{\odot} \le 10^{6}$: $\beta \approx 2
  \pm 0.08$ for Antennae clusters (Zhang \& Fall 1999).
\end{itemize}

\begin{figure}[ht]
\sidebyside
{\centerline{\psfig{file=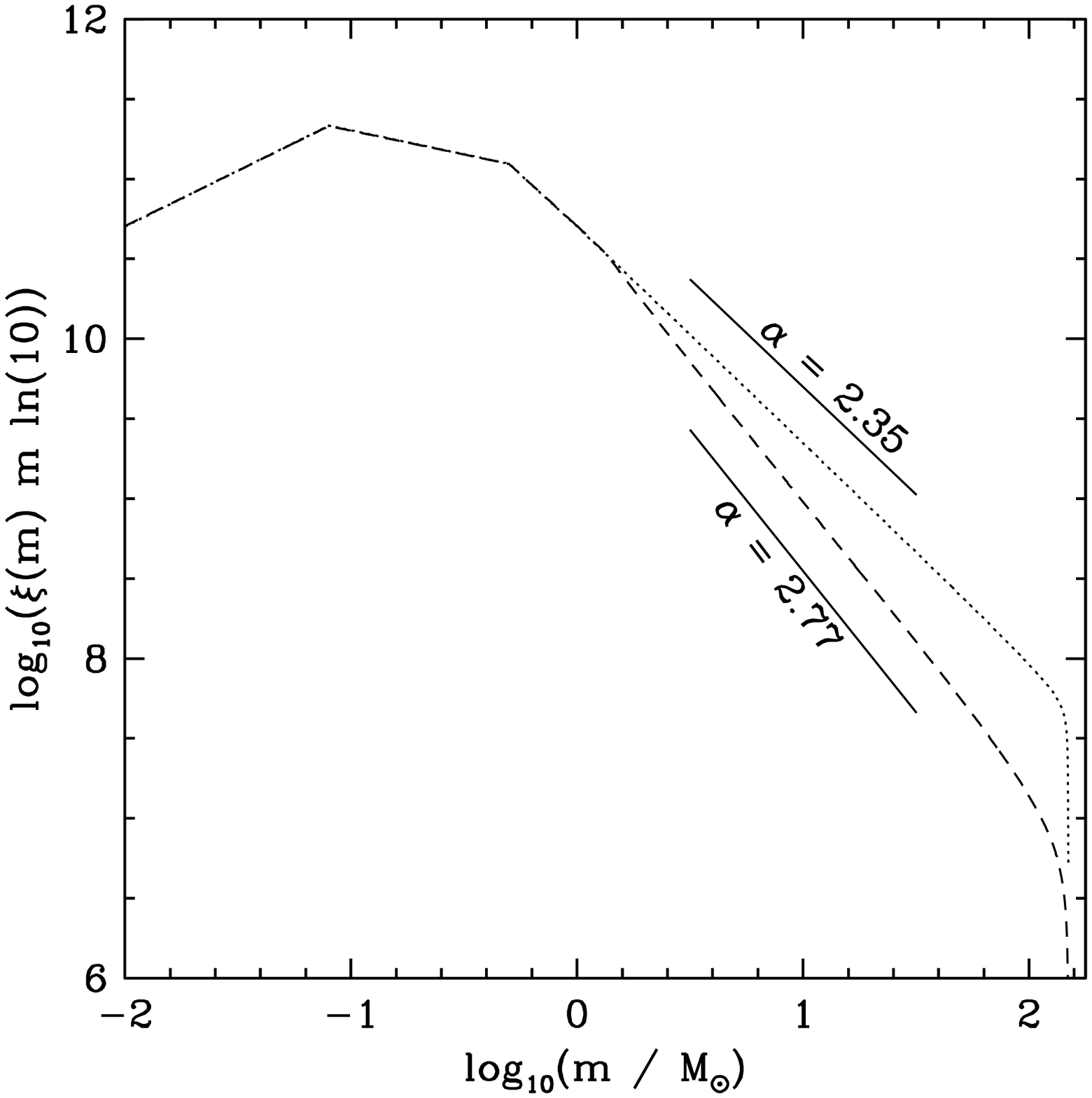,width=6.5cm}}
\vspace*{-1.5cm}
\caption{The dotted line is the canonical stellar IMF, $\xi(m)$, in
  logarithmic units and given by the standard four-part power-law form
  (eq.~\ref{eq:canonicalIMF}). The dashed line is $\xi_{\rm IGIMF}(m)$
  for $\beta=2.2$.  The IMFs are scaled to have the same number of
  objects in the mass interval $0.01-1.0\,M_\odot$.  Note the turn
  down near $m_{\rm max*}=150\,M_\odot$ which comes from taking the
  fundamental upper mass limit explicitly into account (Weidner \&
  Kroupa 2004; 2004a).  Two lines with slopes $\alpha_{\rm line}=2.35$
  and $\alpha_{\rm line}=2.77$ are indicated.}
\label{krFig3}}
{\centerline{\psfig{file=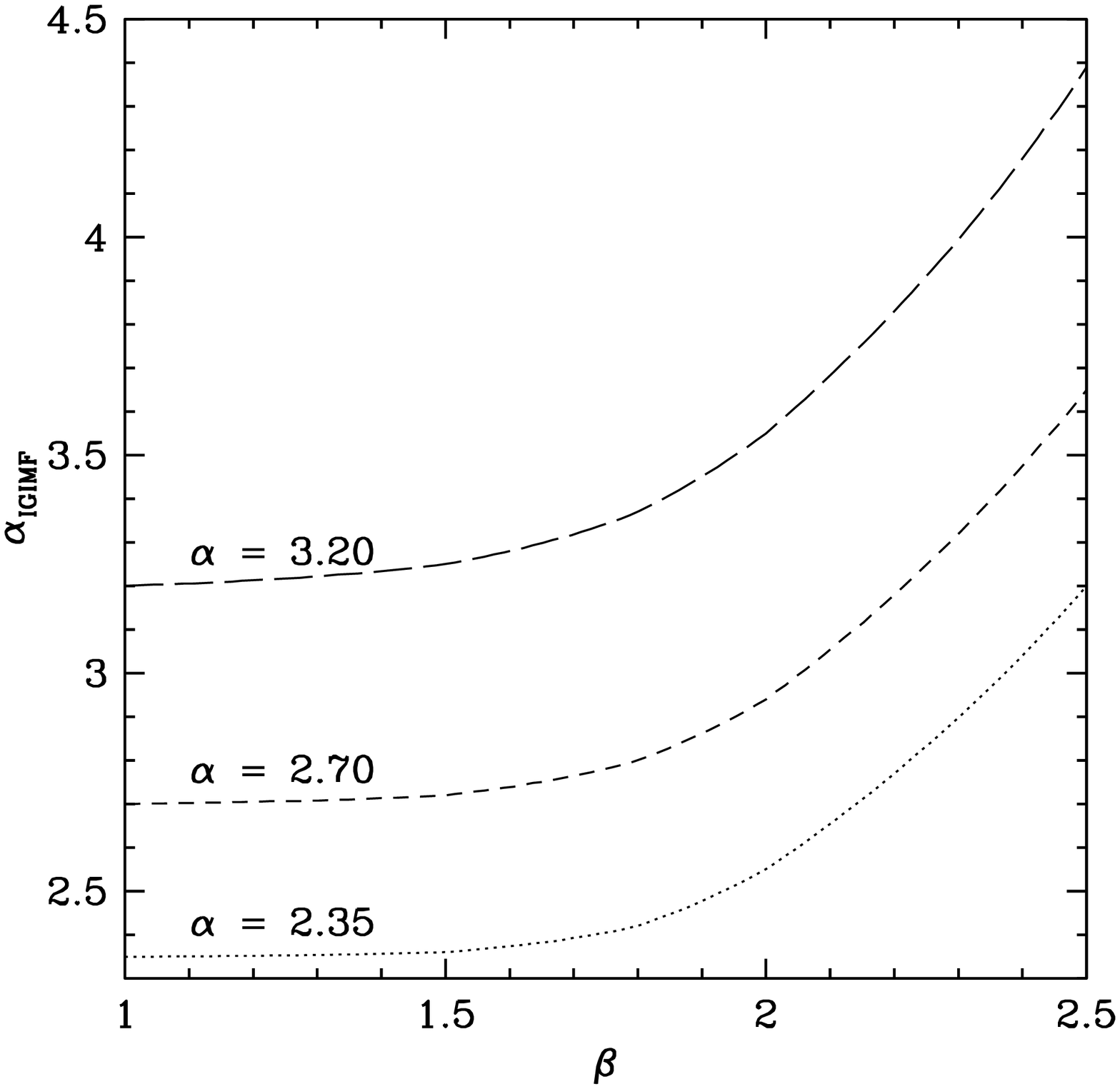,width=6.5cm}}
\vspace*{-1.5cm}
\caption{The IGIMF power-law index $\alpha_{\rm IGIMF}$
($m>1\,M_\odot$) as a function of the star-cluster MF power-law index
$\beta$ for $\alpha=2.35, 2.7, 3.2$.} 
\label{krFig4}}
\end{figure}

\noindent Assuming $\beta \approx 2.2$ and the canonical 
IMF slope $\alpha\equiv \alpha_3 = 2.35$ for stars above
$1\,M_{\odot}$ we get the resulting IGIMF shown in Fig.~\ref{krFig3}
which is considerably steeper than the Salpeter IMF (see Kroupa \&
Weidner 2003 for more details).

In Fig.~\ref{krFig4} the slope of the IGIMF, $\alpha_{\rm IGIMF}$ for
$m>1\,M_\odot$, is studied for various values of $\beta$ and $\alpha$.
The effect is more pronounced the steeper (larger $\beta$) the cluster
mass function is.

\begin{figure}[ht]
{\centerline{\psfig{file=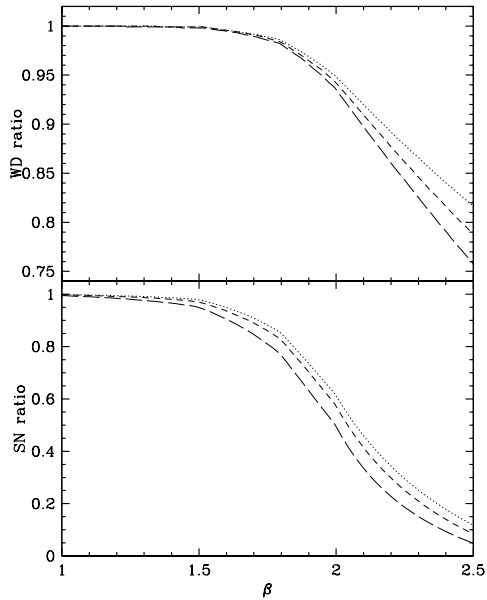,width=6.5cm}}
\vspace*{-0.5cm}
\caption{Upper panel: The ratio of the number of stars in the
  IGIMF (for input stellar IMFs with $\alpha=2.35, 2.7, 3.2$, dotted,
  short-dashed and long-dashed lines, respectively) relative to the
  canonical IMF (Salpeter above $0.5\,M_\odot$) in the mass interval
  $0.8\le m/M_\odot \le 8$ (the relative number of white dwarfs) in
  dependence of the cluster mass-function power-law exponent $\beta$.
  Lower panel: The same except for $8\le m/M_\odot \le m_{\rm
    max*}=150\,M_\odot$ (the relative number of SNII per star).  The
  normalisation of the IMFs is as in Fig.~\ref{krFig3}. Note that the
  panels have different vertical scales.}
\label{krFig5}}
\end{figure}

This has a profound impact on the evolution of galaxies as can be
deduced from Fig.~\ref{krFig5}. In the upper panel is shown the number
of white dwarfs per star relative to the Salpeter IMF. For $\beta$
values above 2 it drops considerably, and in the lower panel, were the
number of supernovae of type~II (SNII) per star is plotted, the effect
is even stronger. For $\alpha=2.35$ and $\beta=2.2$ there are 89\% of
white dwarfs but only 35\% of SNII compared to a Salpeter IMF.

\section{Variations among galaxies}
\label{sec:impli}

\begin{figure}[ht]
\sidebyside
{\centerline{\psfig{file=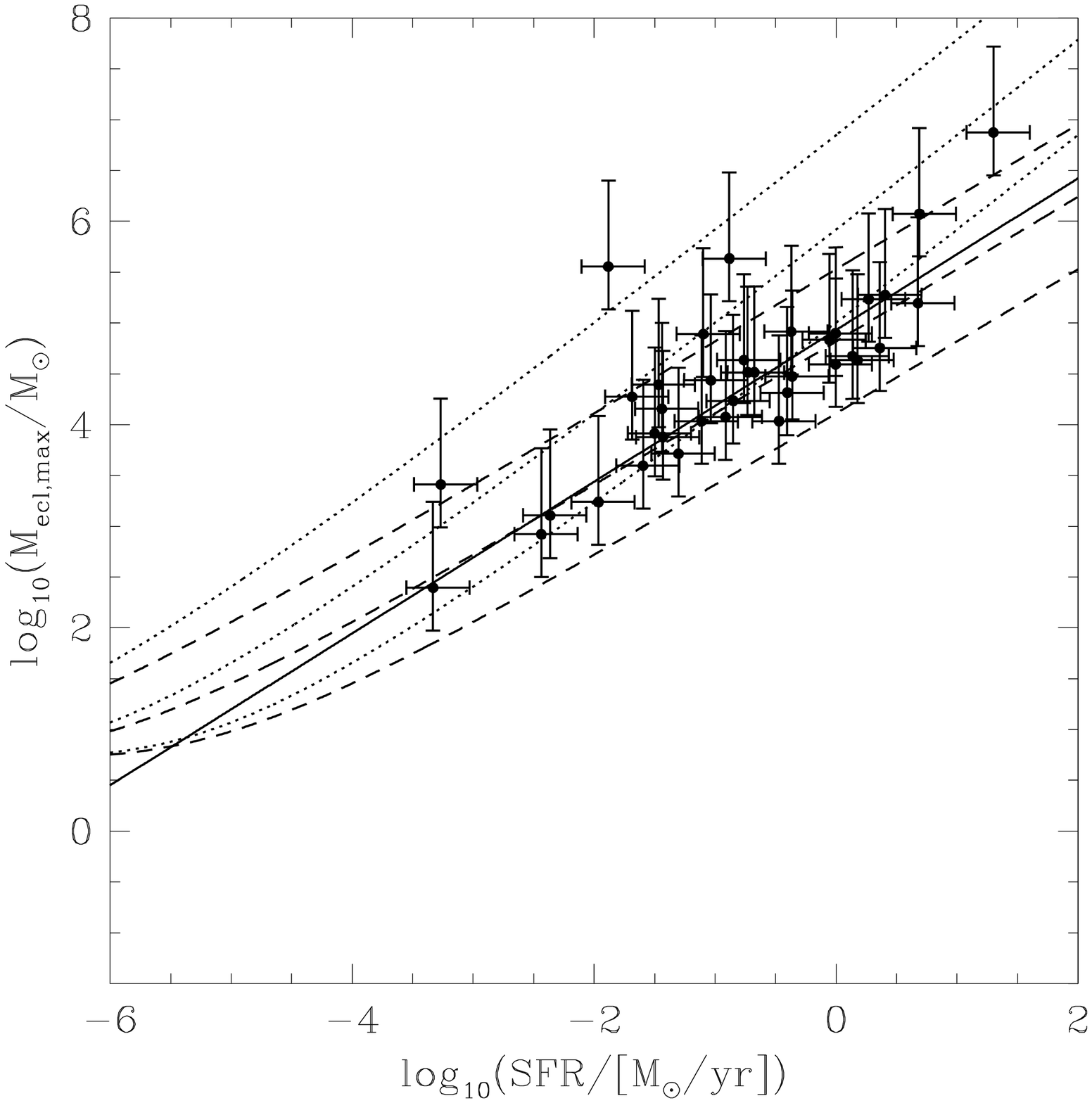,width=6.5cm}}
\vspace*{-1.5cm}
\caption{Maximum cluster mass versus global star-formation rate (SFR),
  both in logarithmic units. Filled dots are observations by Larsen
  (2001; 2002) with error estimates and the linear regression fit is
  the solid line (eq.~\ref{eq:Meclsfr}). The other curves are
  theoretical relations (eq.~\ref{eq:model}) which assume the entire
  young-cluster population forms in $\delta t=1$, 10 and 100~Myr
  (bottom to top). The cluster IMF has $\beta=2$ (dotted curves) or
  $\beta=2.4$ (dashed curves).}
\label{krFig6}}
{\centerline{\psfig{file=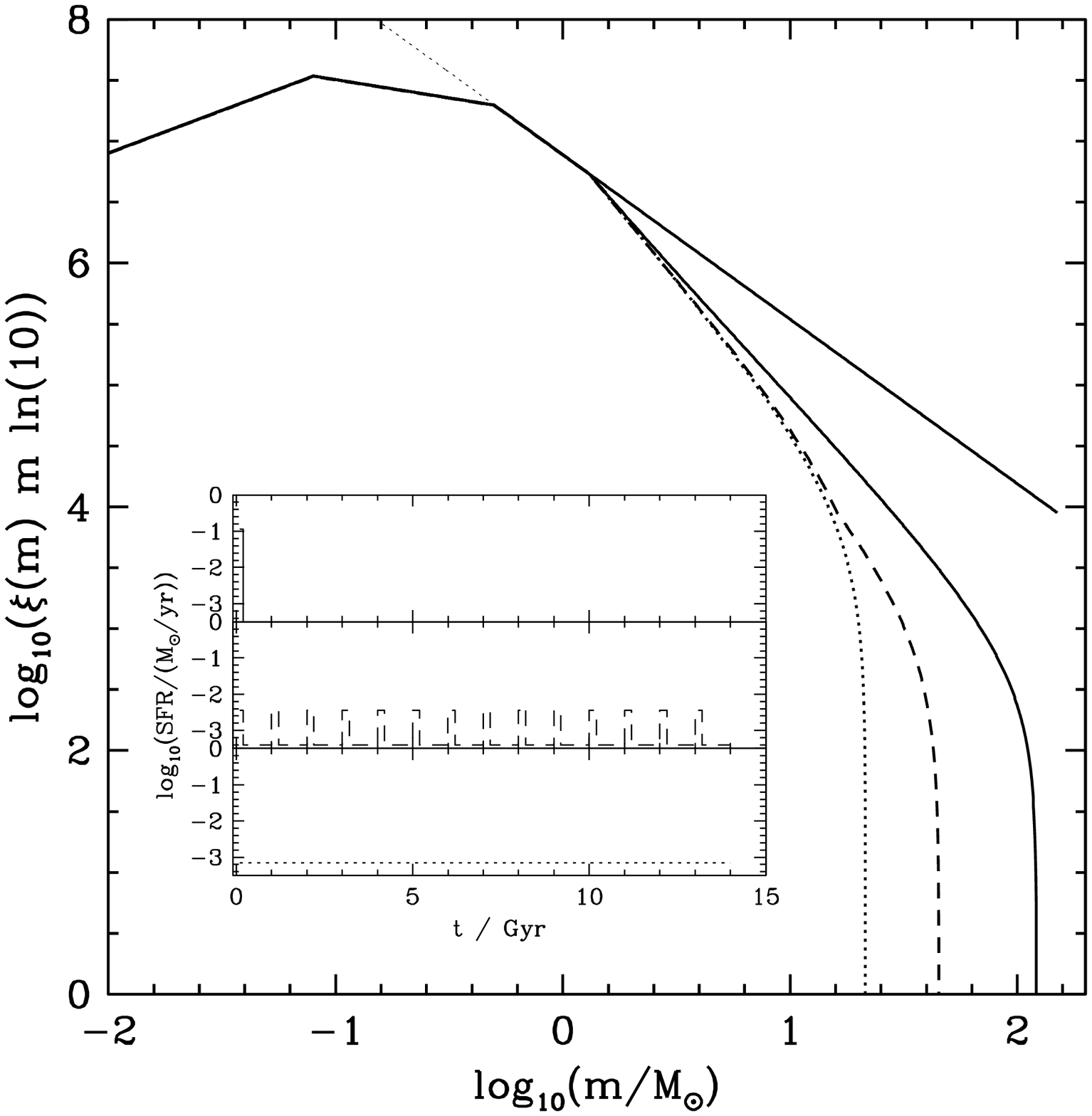,width=6.5cm}}
\vspace*{-1.5cm}
\caption{IGIMFs for three dwarf or low-surface brightness 
  galaxies with a final stellar mass of $10^{7}\,M_{\odot}$ but
  different SFHs. The solid curve results from a single initial
  100~Myr long burst of star formation. The dashed curve assumes a
  periodic SFH with 14~peaks each 100~Myr long and 900~Myr quiescent
  periods in between.  The thick dotted curve assumes a constant SFR
  over 14~Gyr. For all cases the canonical power-law slope above
  $1\,M_{\odot}$ ($\alpha=2.35$) is used and the cluster IMF slope is
  $\beta=2.35$. The straight solid line above $0.5\,M_{\odot}$ shows
  the canonical input IMF for comparison. The thin dotted line is a
  Salpeter IMF extended to low masses. Note the downturn of the IGIMFs
  at high masses. It results from the inclusion of a limiting maximum
  mass, $m_{\rm max}\le m_{\rm max*}$, into our formalism.}
\label{krFig7}}
\end{figure}

\noindent
The star-formation rate (SFR) of a galaxy is given by
\begin{equation}
SFR = M_{\rm tot} / \delta t, 
\label{eq:model}
\end{equation}
where $M_{\rm tot}=\int_{M_{\rm ecl,min}}^{M_{\rm ecl,max}} \, M_{\rm
  ecl}\,\xi_{\rm ecl}(M_{\rm ecl})\,dM_{\rm ecl}$ is the mass in
clusters assembled in time interval $\delta t$.  Lets assume for the
moment that the cluster IMF, $\xi_{\rm ecl}$, is also invariant, and
that $M_{\rm ecl,min}=5\,M_\odot$ is fixed while $M_{\rm ecl,max}$ may
vary. Then the $M_{\rm ecl,max}$ vs $SFR$ relation can be calculated
for different $\delta t$. A comparison of this simple theory with
empirical data is provided in Fig.~\ref{krFig7}. The figure plots
empirical maximum star-cluster masses vs galaxial global SFRs, and a
fit to these data yields
{\small
\begin{equation}
\label{eq:Meclsfr}
\log_{10}(M_{\rm eclmax}) = \log_{10}(k_{\rm ML}) + (0.75(\pm
0.03) \cdot \log_{10}{SFR}) + 6.77(\pm 0.02),
\end{equation}}
\noindent
where $k_{\rm ML}$ is the mass-to-light ratio. Fig.~\ref{krFig6} shows
this relation. Interestingly, the theory is nicely consistent with the
data for $\beta\approx 2.2$ and $\delta t\approx 10$~Myr, which
suggests that the cluster IMF may indeed be quite invariant, and that
a galaxy is able to assemble complete star-cluster systems within
typically 10~Myr independent of the SFR (Weidner, Kroupa \& Larsen
2004).

With the use of eq.~\ref{eq:Meclsfr} the time-dependent IGIMF of
galaxies of different types can be calculated: On specifying a SFR,
$M_{\rm ecl,max}$ follows (eq.~\ref{eq:Meclsfr}) and from
eq.~\ref{eq:igimf} the galaxial IMF can be computed in dependence of
the star-formation history by adding up all galaxial IMFs generated in
each multiple-$\delta t$-epoch until the present. As a
straight-forward result we expect the galaxial IMF to be steeper
(larger $\alpha_{\rm IGIMF}$) for low-mass galaxies than for massive
galaxies because the average SFR is lower for the former. We also
expect a dependence of the galaxial IMF on the star-formation history
(SFH) of a galaxy.

\begin{figure}[ht]
\sidebyside
{\centerline{\psfig{file=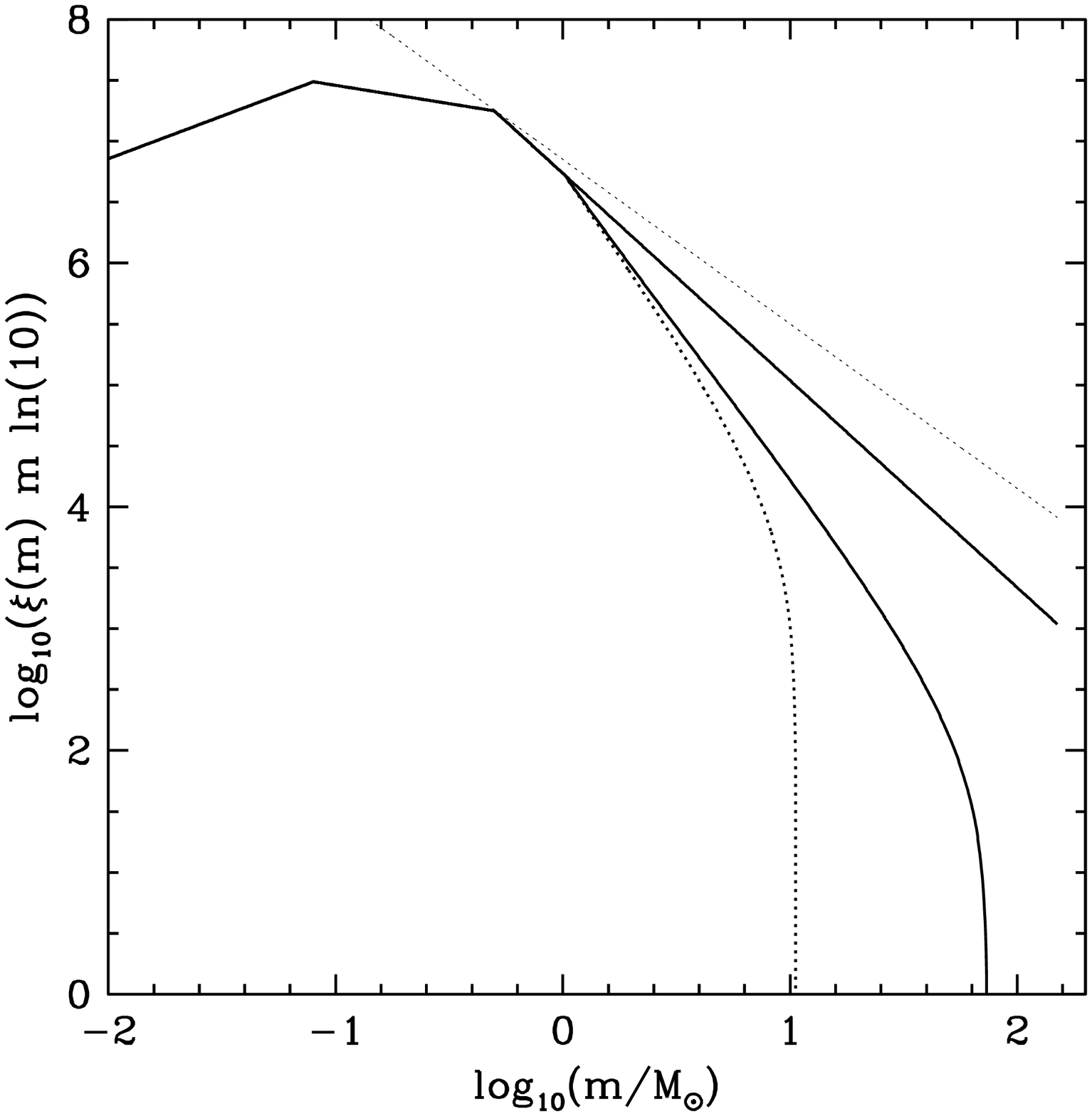,width=6.5cm}}
\vspace*{-1.5cm}
\caption{As Fig.~\ref{krFig7} but only for
  two cases of the SFH: the burst case (solid line) and the low
  continuous case (thick dotted line), assuming $\alpha=2.70$.} 
\label{krFig8}}
{\centerline{\psfig{file=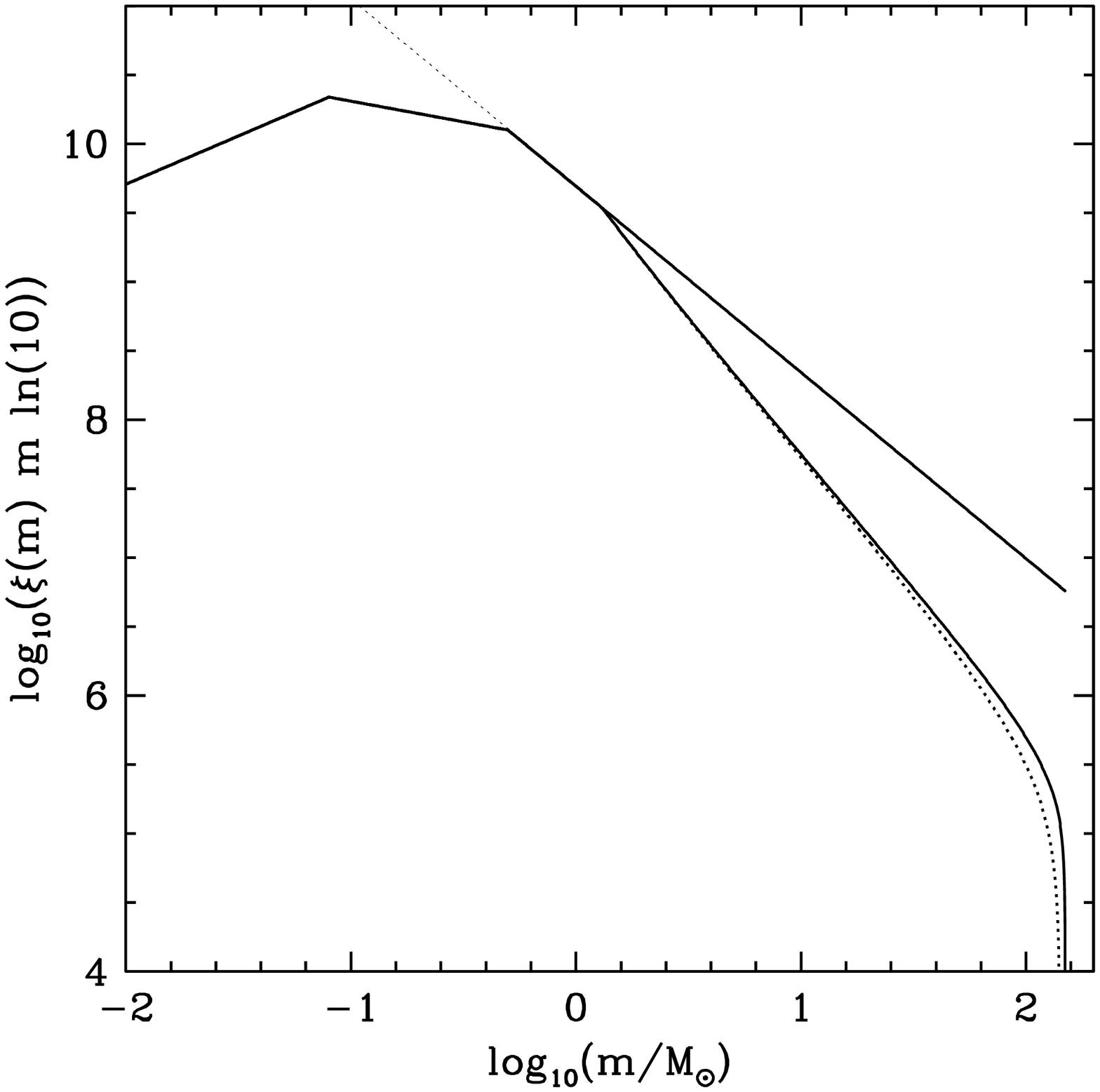,width=6.5cm}}
\vspace*{-1.5cm}
\caption{As Fig.~\ref{krFig7} but for a galaxy with stellar mass 
  $10^{10}\,M_{\odot}$ and canonical $\alpha=2.35$. The IGIMF is
  only shown for a burst SFH (solid curve) and for a continuous
  SFH (dotted curve).} 
\label{krFig9}}
\end{figure}

\begin{figure}[ht]
{\centerline{\psfig{file=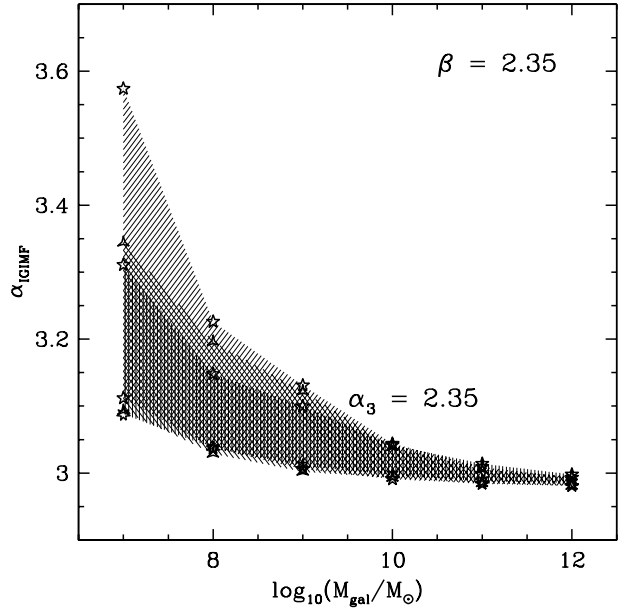,width=8.5cm}}
\vspace*{-2cm}
\caption{The resulting IGIMF slopes above $1\,M_{\odot}$ are
  shown in dependence of stellar galaxy mass for different models, as
  indicated.  The lower bounds of the shaded areas are for an initial
  star formation burst forming the entire stellar galaxy while the
  upper bounds are derived for a continuous SFR.  The various symbols
  correspond to calculated models. Symbols lying within the shades
  area correspond to models with an episodic SFH.}
\label{krFig10}}
\end{figure}

These expectations are born out. The final IGIMF of a dwarf or
low-surface brightness galaxy with a stellar mass of $M_{\rm gal} =
10^{7} M_{\odot}$ assuming a stellar IMF slope $\alpha = 2.35$ is
shown in Fig.~\ref{krFig7}, and in Fig.~\ref{krFig8} for $\alpha =
2.70$.  In the case with $\alpha = 2.35$ three SFHs are considered. A
single burst of star-formation followed by no further formation (solid
line), an episodic SFR with~100 Myr long peaks every 900~Myr
(long-dashed line) and a constant SFR over 14~Gyr (dotted line). The
influence of the SFH on the IGIMF is significant for galaxies with a
small mass in stars. For a steeper input IMF (Fig.~\ref{krFig8}) this
effect is even more pronounced.  Such a steep IMF slope may possibly
be the true value if the observationally derived stellar IMFs for
massive stars are corrected for unresolved binaries (Sagar \& Richtler
1991). In galaxies with a large stellar mass, $M_{\rm gal} = 10^{10}
M_{\odot}$ (Fig.~\ref{krFig9}), the sensitivity of $\alpha_{\rm
  IGIMF}$ on the SFH becomes negligible because the average SFR is
always large enough to sample the star-cluster IMF to the fundamental
stellar upper mass limit. In summary, the IGIMF slope, $\alpha_{\rm
  IGIMF}$, is shown as a function of $M_{\rm gal}$ in
Fig.~\ref{krFig10}.  The differently shaded regions are for single
burst, episodic and a constant SFR, respectively.

What implications beyond a reduction of the number of SNII do these
findings have?  The observed diversity of metallicities in different
dwarf galaxies of rather similar mass (Mateo 1998; Garnett 2004) may
be explained by this effect without invoking individually fitted
effective yields as are needed in chemo-dynamical models with
'standard' IMFs (Lanfranchi \& Matteucci 2004). On the other hand, our
models indicate that massive spirals and ellipticals should show less
variations in the IGIMF and thus metallicity due to their high average
SFR which is actually what is observed in nearby galaxies (fig.~9 in
Garnett 2004).  Current available models for the chemical evolution of
galaxies have difficulties to reproduce the chemical abundances in
disc galaxies with a standard galaxial Salpeter IMF, as it produces
too many metals (Portinari et al. 2004). This problem may be
resolvable by the model presented here but detailed chemo-dynamical
calculations incorporating our approach need to be computed. Finally,
the mass-to-light ratia of galaxies are too low for models that assume
invariant galaxial Salpeter IMFs; the real $M/L$ ratia are larger.

\section{Conclusions}
\label{sec:concs}
\begin{itemize}

\item[1.] IGIMFs are steeper than the stellar IMF.

\item[2.] Therefore there are significantly fewer SNII per G-type star per
  galaxy.

\item[3.] Chemical enrichment and $M/L$ ratios of galaxies calculated 
with an invariant Salpeter IMF are wrong.

\item[3.] The IGIMFs vary from galaxy to galaxy.

\item[4.] Dwarf galaxies show the largest variations: $3.1 \simless
  \alpha_{\rm IGIMF} \simless 3.6$~for $\alpha_{3, \rm true} = 2.35$
  (Salpeter) and $\beta=2.35$.

\item[5.] The sensitivity of the IGIMF on the SFH increases with
  $\alpha_{3}$. The true value of $\alpha_{3, \rm true}$ for the
  stellar IMF may actually be significantly larger than~2.35\,:

 \item[6.] Understanding the corrections for unresolved multiples on
  the derivation of the high-mass part of the stellar IMF is of
  fundamental importance!

\end{itemize}

\vspace{3mm}

\noindent{\bf Acknowledgement:}\\ This work is being supported 
by DFG grant KR1635/3-1. 

%

\end{document}